\documentclass[aps,twocolumn,superscriptaddress]{revtex4-1}
\pdfoutput=1

\usepackage{graphicx}
\usepackage{amsmath}
\usepackage{amssymb}
\usepackage{color}
\usepackage{bm}
\usepackage{subfigure}
\usepackage{appendix}

\definecolor{red}{rgb}{0.75,0,0}
\definecolor{blue}{rgb}{0,0,0.75}
\definecolor{green}{rgb}{0,0.5,0}

\begin{document}
\title{Optimal shapes and stresses of adherent cells on patterned substrates}

\author{Shiladitya Banerjee}
\affiliation{James Franck Institute, The University of Chicago, Chicago, Illinois 60637, USA}
\author{Rastko Sknepnek}
\affiliation{Department of Physics and Syracuse Biomaterials Institute, Syracuse University, Syracuse, New York 13244, USA}
\affiliation{School of Engineering, Physics, and Mathematics, University of Dundee, Dundee DD1 4HN, UK}
\author{M. Cristina Marchetti}
\affiliation{Department of Physics and Syracuse Biomaterials Institute, Syracuse University, Syracuse, New York 13244, USA}

\begin{abstract}
We investigate a continuum mechanical model for an adherent cell on two dimensional adhesive micropatterned substrates. The cell is modeled as an isotropic and homogeneous elastic material subject to uniform internal contractile stresses. The build-up of tension from cortical actin bundles at the cell periphery is incorporated  by introducing an energy cost for  bending of the cell boundary, resulting to a resistance to changes in local curvature. Integrin-based adhesions are modeled as harmonic springs, that pin the cell to adhesive patches of a predefined geometry. Using Monte Carlo simulations and analytical techniques we investigate the competing effects of bulk contractility and cortical bending rigidity in regulating cell shapes on non-adherent regions. We show that the crossover  from convex  to concave cell edges is controlled by the interplay between contractile stresses and boundary bending rigidity. In particular, the cell boundary becomes concave beyond a critical value of the contractile stress that is proportional to the cortical bending rigidity. Furthermore, the intracellular stresses are found largely concentrated at the concave edge of the cell. The model can be used to generate a cell-shape phase diagram for each specific adhesion geometry. 
\end{abstract}

\maketitle

\section{Introduction}
\label{sec:Intro}
Living cells actively probe physical cues in their  environment via receptor-ligand adhesion complexes that link the actomyosin cytoskeleton to the extracellular matrix (ECM)~\cite{Schwarz2012}. The cellular microenvironment, comprising of the ECM and of neighboring cells, imposes specific boundary conditions that can regulate physiological processes such as cell differentiation, division and motility, as well as cell architecture and polarity~\cite{Discher2005}.  Myosin motors generate contractile stresses in the actin cytoskeleton that are transmitted to the substrate  by focal adhesions. The traction stresses exerted by the cells on the substrate are thus very sensitive to the stiffness of the substrate as well as to the adhesion geometry. Cell morphology in turn is directly affected by traction stresses through the tension that builds up in the actomyosin stress fibers. It has been shown that the substrate stiffness plays a crucial role in regulating the cell spread area, the magnitude of traction 
forces and the cell morphology~\cite{Yeung2005,Lo2000,Ghibaudo2008,Chopra2011}. Much less explored is the role of adhesion geometry in regulating the spatial distribution of cellular stresses. Micropatterning has emerged as a powerful tool  to investigate the interplay of mechanics and cytoskeletal architecture in controlling cell morphology  by specific tuning of the geometry of the adhesion sites~\cite{Thery2010}. When plated on small micropatterns, cells are unable to grow, thus showing high apoptotic rate~\cite{Chen1997}. Large adhesive patches, in contrast, favor cell spreading and promote the assembly of contractile stress fibers along the cell's perimeter~\cite{Thery2006}. These peripheral stress fibers  interconnect focal adhesions and yield concave arcs of constant curvature in the nonadherent portions of the cell boundaries. In addition, traction forces tend to localize in regions of high curvature at the boundary~\cite{Roca2008,Rape2011}. The model proposed here allows to separately study the roles of cell 
contractility and mechanical properties of peripheral cell fibers in controlling cell shape. Future comparison with experiments where both quantities can be 
  perturbed using pharmacological interventions~\cite{Barziv1999,Thery2006} may provide a quantitative understanding of the relative importance of boundary and bulk properties in determining 
steady state cell shapes.

Various successful theoretical models of single and multi-cell mechanics have been proposed over the past decade that address the role of ECM elasticity in regulating cell behavior~\cite{Schwarz2013}. Previous work has addressed the interplay between cell mechanics and geometry by either focusing solely on the elasticity of the cell boundary~\cite{Bischofs2009,Banerjee2013b} or by considering only the bulk of the cell, described via continuum mechanics~\cite{Banerjee2011,Edwards2011,Pathak2008,Banerjee2013a}, by a cellular Potts model~\cite{Vianay2010}, or as a polymer network~\cite{Torres2012}. These models highlight the competing roles of cell contractility and substrate stiffness in regulating polymorphic cell shapes.

\begin{figure}
\begin{centering}
\includegraphics[width=1\columnwidth]{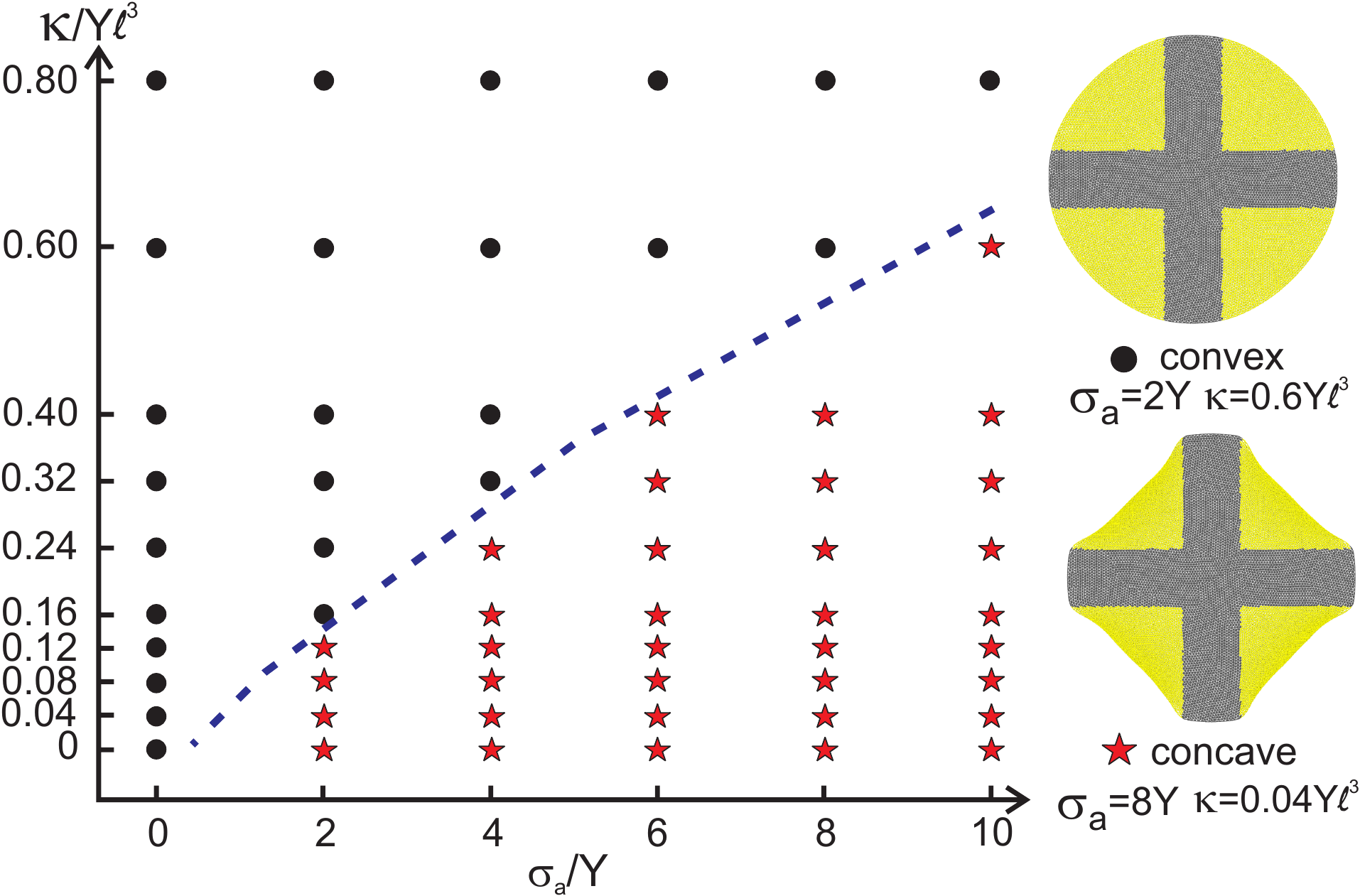}
\par\end{centering}
\caption{Shape phase diagram for a cell adhering to a cross shaped micropattern (grey region of the cell images) as a function of the bending rigidity $\kappa$ of the cell boundary in units of $Y\ell^3$ and the bulk contractile stress $\sigma_{a}$ in units of $Y$, where $Y$ is the cellular Young's modulus and the $\ell$ is the interparticle distance in the triangulation. The blue dashed line is a guide to the eye.
\label{fig:phase-diagram}}
\end{figure}

Continuum models of cell mechanics have assumed that the material constants describing the cell are spatially homogeneous. Cell material properties are, however, highly heterogeneous. In particular, experiments have shown strong differences in the mechanical properties of the bulk and boundary 
regions of the cell~\cite{Heidemann2004}. Increased tension and rigidity of cell boundaries can spontaneously arise during adhesion as a result of the assembly of peripheral stress fibers consisting of thin bundles of semiflexible actin filaments. Due to thermal and active forces these bundles considerably bend generating non-uniform peripheral tensions. Cell boundary can also resist changes in local curvature due to contact forces at the three-phase contact line between the cell, the substrate and the ambient medium. Motivated by these observations, in this paper we  couple cell contour elasticity~\cite{Bischofs2009,Banerjee2013b} to a continuum description of bulk cell mechanics~\cite{Banerjee2011,Edwards2011} to investigate  the cooperative roles of cortical elasticity, bulk elasticity and active contractility in controlling cell shapes on non-uniform  adhesion patterns. Non-uniform tension and elasticity is incorporated in the model by introducing a penalty for bending deformations of the cell periphery~\cite{Banerjee2013b}.  Using a combination of Monte Carlo simulation and analytical studies, we examine the interplay of bulk contractility and cortical tension in controlling morphological transitions in adherent cells and propose a cell shape phase diagram for specific adhesion geometries. An example of such a phase diagram for a cross-shaped adhesion pattern is shown in Fig.~\ref{fig:phase-diagram}. See section~\ref{sub:Optimal_shapes} for details.

The paper is organized as follows. In section \ref{sec:Cont_model}, we describe a continuum mechanical model for a thin adherent cell as an isotropic and homogeneous elastic material, subject to a homogeneous negative pressure, embodying active contractility. Cell-ECM adhesions are modeled as linear springs distributed non-uniformly along the cell-substrate interface. Cortical tension is described via a penalty for bending deformations of the cell periphery. In Section \ref{sec:Num_sim} we discuss the steady shapes of  cells adherent to different concave micropatterns obtained via  Monte Carlo simulations (numerical details are given in the Appendix). Our simulations suggest a transition between convex and concave morphologies as a function of cell contractility and bending rigidity, that is captured by an analytically solvable model for the cell boundary presented in Section \ref{sec:Boundary_shapes}. We conclude with a brief discussion.

\vspace{1in}

\section{Continuum Mechanical Model}
\label{sec:Cont_model}
We consider the mechanical equilibrium of a stationary cell strongly adherent to a soft elastic substrate. We assume that the cell's interior can be described as an isotropic homogeneous elastic material and neglect all dissipation. We further  neglect all out-of-plane deformations of the cell and assume that its thickness is uniform throughout its entire area and remains unaffected by the substrate-induced 
deformations. The bulk elastic  energy of our model cell is given by
\begin{eqnarray}
E_{el} & = & \int_{A_{0}}dA\frac{Eh}{2\left(1+\nu\right)}\left(\frac{\nu}{1-\nu}u_{\gamma\gamma}^2+u_{\alpha\beta}^{2}\right)\;\label{eq:el_eng}
\end{eqnarray}
with $E$ the three-dimensional Young's modulus, $h$ the cell
thickness, $\nu$  the Poisson's ratio, and $u_{\alpha\beta}=\frac{1}{2}\left(\partial_{\alpha}u_{\beta}+\partial_{\beta}u_{\alpha}+\partial_{\alpha}u_{\gamma}\partial_{\beta}u_{\gamma}\right)$
($\alpha,\beta,\gamma\in\left\{ x,y\right\} $) the strain tensor. We retain the nonlinearity in strain tensor to allow for the possibility of large strains that can arise even for small displacements~\cite{audoly2010}. The nonlinear terms essentially describe strain stiffening which is indeed expected in crosslinked actin networks~\cite{Gardel2004}.
The two-dimensional displacement vector $\vec{u}$ is 
defined as $\vec{u}\left(\vec{r}_{0}\right)=\vec{r}-\vec{r}_{0}$,
where $\vec{r}_{0}$($\vec{r}$) is a material point before (after)
the deformation. The integral is calculated over the area $A_0$ of the undeformed (reference)
state and  summation over pairs of repeated indices is assumed.  
Cell contractility arising from myosin motors is modeled
as a homogeneous negative pressure resulting in an additional contribution to the cell's energy, given by
\begin{equation}
E_{active}=\sigma_{a}\int_{A_{0}}dA\ u_{\gamma\gamma}\;,\label{eq:active}
\end{equation}
where $\sigma_{a}>0$ is a parameter controlling active  contractility, determine by concentration of myosin motors and rate of ATP consumption. At the continuum scale, it controls active contributions to the cell's surface tension~\cite{Mertz2012}. 
Cell adhesion to the
substrate is modeled via a harmonic potential with a position-dependent
rigidity parameter $\Gamma(\vec{r})$
\begin{equation}
E_{adh}=\frac{1}{2}\int dA\ \Gamma\left(\vec{r}\right)\left|\vec{r}-\vec{r}_{a}\right|^{2}\;,\label{eq:adh}
\end{equation}
with $\vec{r}_{a}$  the position of focal adhesions on the substrate. The rigidity parameter $\Gamma$ is nonzero over the  adhesion region and zero elsewhere. Thus we allow for non-uniformity in the geometry of cell-substrate adhesions as can be realized experimentally using micropatterning techniques~\cite{Thery2006}. The assumption of \emph{local} adhesive interactions with the underlying substrate strictly holds for elastic substrates that are much thinner than the cell perimeter or on soft microposts~\cite{Banerjee2012}. The rigidity parameter $\Gamma$ depends on the elastic modulus of the underlying substrate as well as on the stiffness $k_f$ of focal adhesions. For an elastic substrate of shear modulus $\mu_s$ and thickness $h_s$, with focal adhesion density $\rho_f$, $\Gamma$ is given by, $\Gamma^{-1}=\left(k_f \rho_f\right)^{-1} + \left(\mu_s/h_s\right)^{-1}$. Traction force density is therefore given by, $\vec{T}=\frac{1}{h}\delta E_\text{adh}/\delta\vec{u}=\frac{1}{h}\Gamma(\vec{r})(\vec{r}-\vec{r}_a)$.
Finally, we assign a bending penalty to the cell's perimeter, reflecting the resistance of cortical actin bundles to changes in curvature,
\begin{equation}
E_{bend}=\kappa\oint ds\ c^{2},\label{eq:bending}
\end{equation}
where $\kappa$ is the bending rigidity, $c=\left|\gamma''\left(s\right)\right|$
is the curvature of the boundary, with $\gamma\left(s\right)$  a
parametric curve describing the cell boundary, and the line integral is calculated
along the cell boundary. 

The optimal shape of the cell is obtained by minimizing the total mechanical energy $E$, that is given as the sum of elastic, active, adhesion, and boundary bending energies, $E=E_{el}+E_{active}+E_{adh}+E_{bend}$.

\begin{figure}
\begin{centering}
\includegraphics[width=0.8\columnwidth]{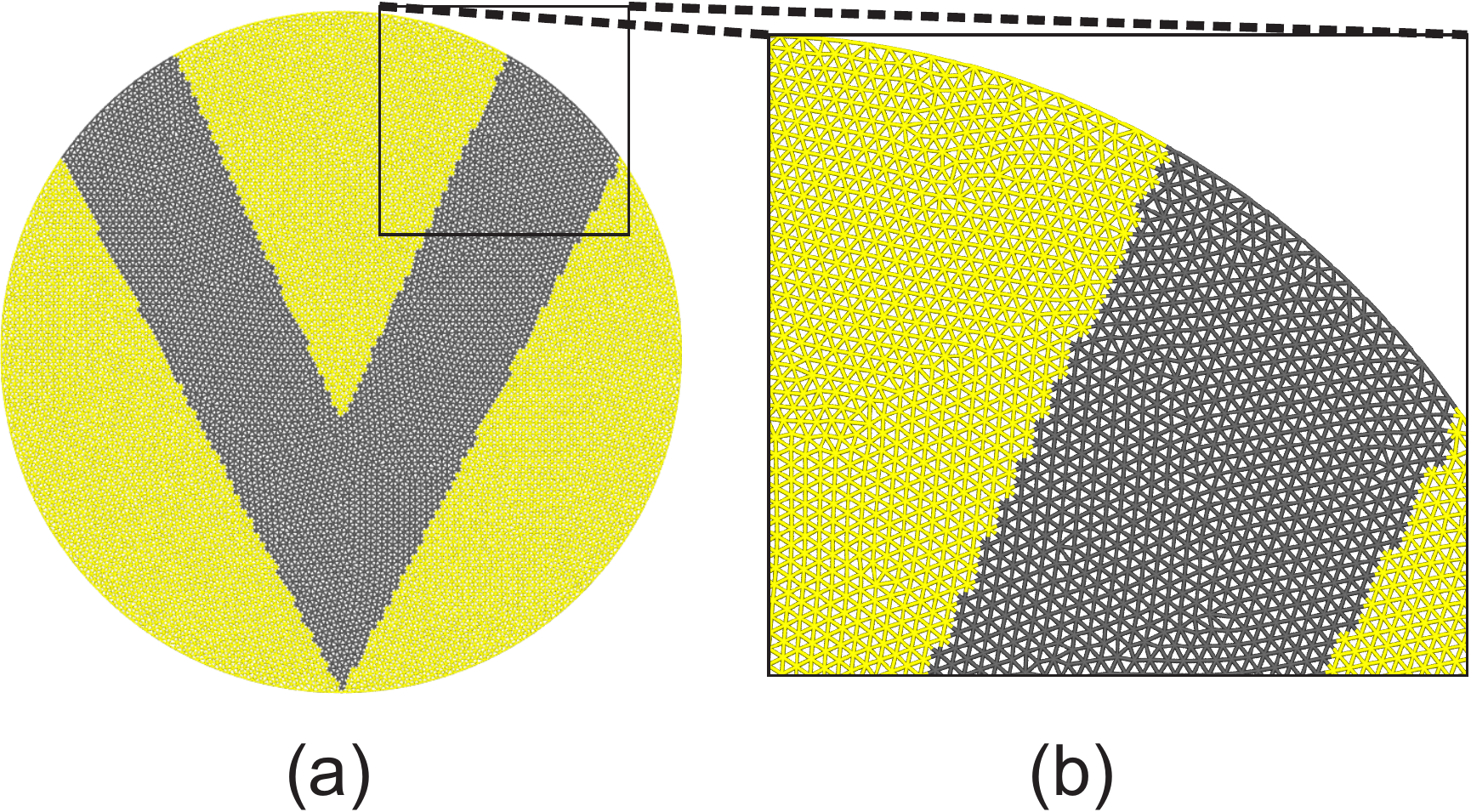}
\par\end{centering}
\caption{(a) Initial configuration of the "V-shape" adhesion pattern. The adhesive region where the cell is strongly anchored to the V-shaped micro-pattern on the substrate ($\Gamma\not=0$) is indicated in grey, whereas the non-adherent portion of the cell ($\Gamma=0$)  is indicated in yellow.  (b) Zoom-in of the upper right corner, showing the triangulation.}
\label{fig:intitial}
\end{figure}
\section{Numerical Simulations}
\label{sec:Num_sim}
The minimal energy cell shapes have been determined numerically by a Monte Carlo study of a discrete representation of the continuum model introduced in Section \ref{sec:Cont_model}.
The discrete representation of the undeformed cell is a triangulated disk.
The initial configuration is built by randomly placing $N\approx10^{4}$
particles on a disk of radius $R_0$. Particles are assumed to interact pairwise via a Weeks-Chandler-Andersen potential~\cite{Weeks1971}
and their positions are equilibrated
using a standard Monte Carlo simulation with canonical ($NVT$) Metropolis algorithm. A typical equilibrated configuration
is stored and the positions of the particles of that configuration
are then used as nodes to construct a Delaunay triangulation. The resulting triangulation for a V-pattern is shown in Fig.~\ref{fig:intitial}. 
We note that the initial density of points in the disk is chosen such
that even in the equilibrium state there is a substantial overlap
between neighbors, thus ensuring a densely packed distribution
of points. As a result the equilibrium distribution of the interparticle distances is rather narrow and its mean, denoted as $\ell$, represents a suitable unit of length. In the following, all distances are measured in units of $\ell$ and all energies are measured in units of $Y\ell^{2}$, where $Y=Eh$ is the two-dimensional Young's modulus. The substrate rigidity $\Gamma$ has units of $Y\ell^{2}$ and the bending rigidity of cortical stress fibers has units of $Y\ell^{3}$. The low energy configurations are obtained using simulated annealing Monte Carlo (see Appendix for details).

\subsection{Optimal shapes}
\begin{figure*}[htb]
\includegraphics[scale=0.9]{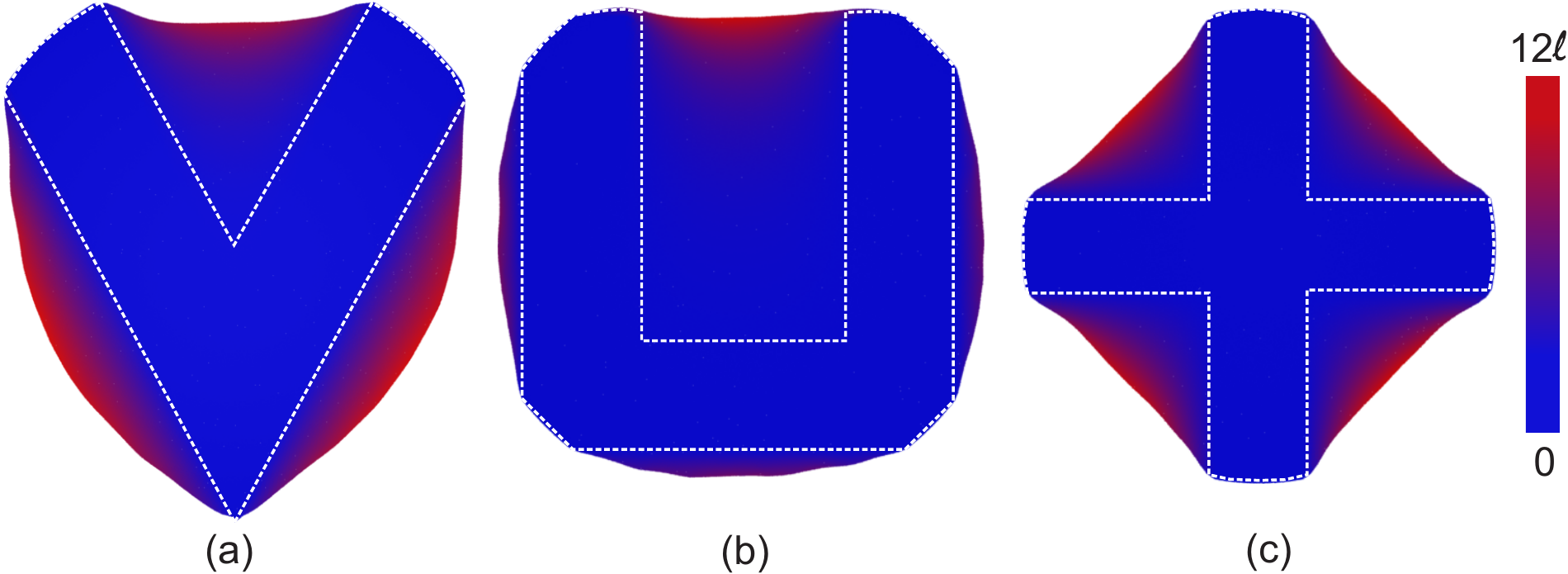}
\caption{Relaxed shapes of (a) $V$-pattern with $\kappa=0.4\ Y\ell^{3}$, (b) $U$-pattern with $\kappa=0.08\ Y\ell^{3}$, and (c) cross-pattern with $\kappa=0.4\ Y\ell^{3}$ and $\sigma_{a}=100Y$. The white dashed lines indicate the boundary of the micropattern: the cell is anchored  inside this region and is free to contract outside. The color scale shows the distribution of the displacements with respect to the reference circular configuration.}
\label{fig:shapes}
\end{figure*}
\label{sub:Optimal_shapes}
We performed a series of simulations for $V$, $U$, and cross shaped micro-patterns, corresponding to the white dashed outlines shown in Fig.~\ref{fig:shapes}. 
The relaxed shapes for three non-convex
patterns ($U$, $V$, and cross) obtained for a fixed value of the rigidity of adhesions $\Gamma_{grey}=10^6$, and $\nu=1/3$ are shown in Fig.~\ref{fig:shapes}. In all cases, the non-adherent cell edges spanning two pinning regions are clearly concave. The relaxed shapes can be qualitatively compared with experiments on concave micropatterns~\cite{Thery2006}.
Our simulations suggest that there
is a transition between the concave and convex morphologies as
a function of $\sigma_{a}$ and $\kappa$. In Figure~\ref{fig:phase-diagram}, we show a sample cell shape phase diagram  as a function of the cortical bending rigidity $\kappa$ and the active contractile stress $\sigma_{a}$ for the cross shaped micropattern. The figure indicates that the cell boundary is concave at high values of the contractile stress $\sigma_a$, whereas convexity is ensured at high values of bending rigidity. The phase boundary between convex and concave shapes appears to be linear in the $\sigma_a -\kappa$ plane. To justify this observation, in the next section  we study analytically the shape of  the cell boundary in the non-adhesive regions, considering small deformation about a circular configuration.

\subsection{Optimal Stresses}
\label{sub:Optimal_stress}
\begin{figure*}[htb]
\begin{minipage}[t]{0.32\textwidth}%
\begin{center}
\includegraphics[scale=0.22]{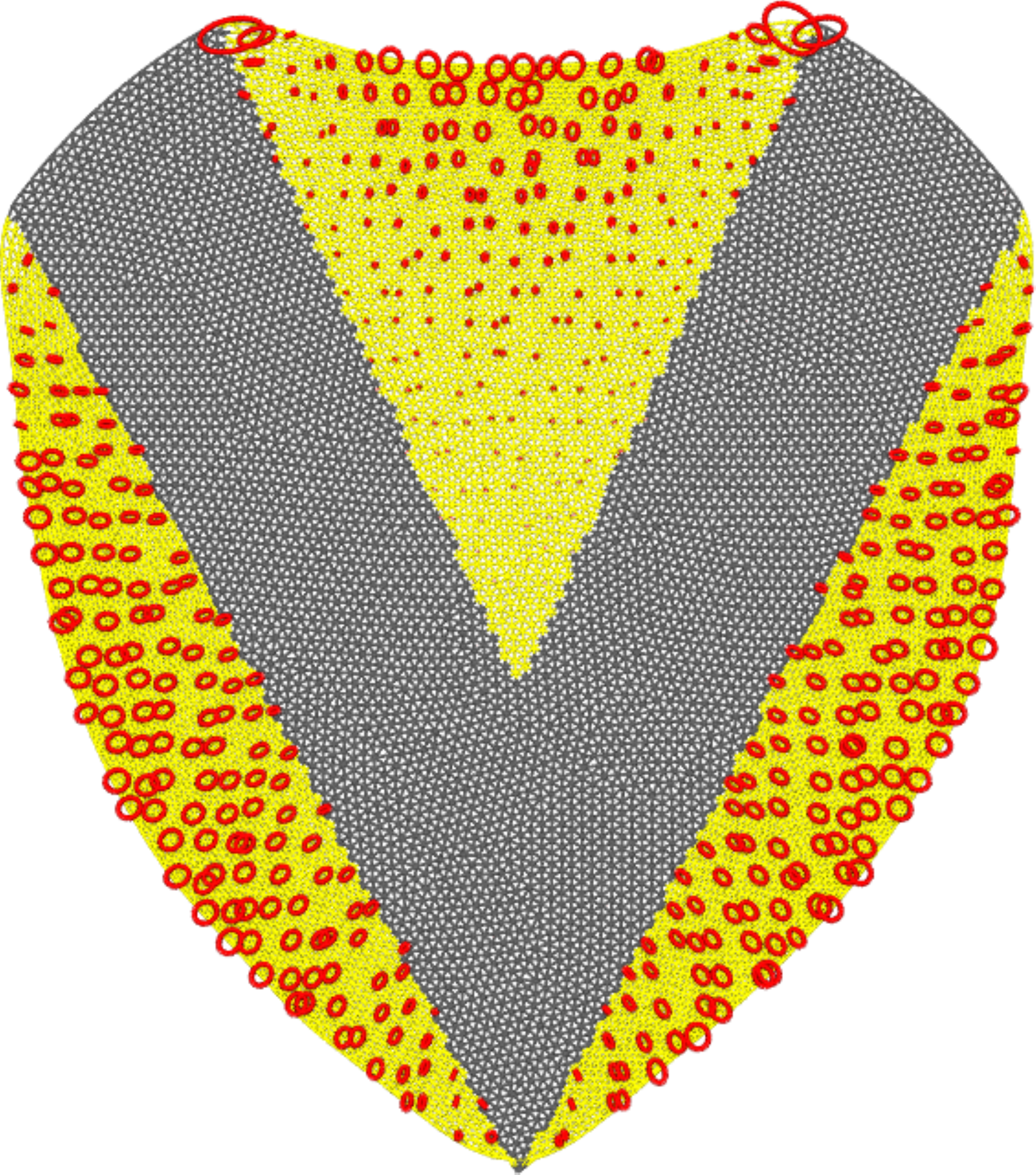}
\par
(a)
\end{center}%
\end{minipage}%
\begin{minipage}[t]{0.32\textwidth}%
\begin{center}
\includegraphics[scale=0.2]{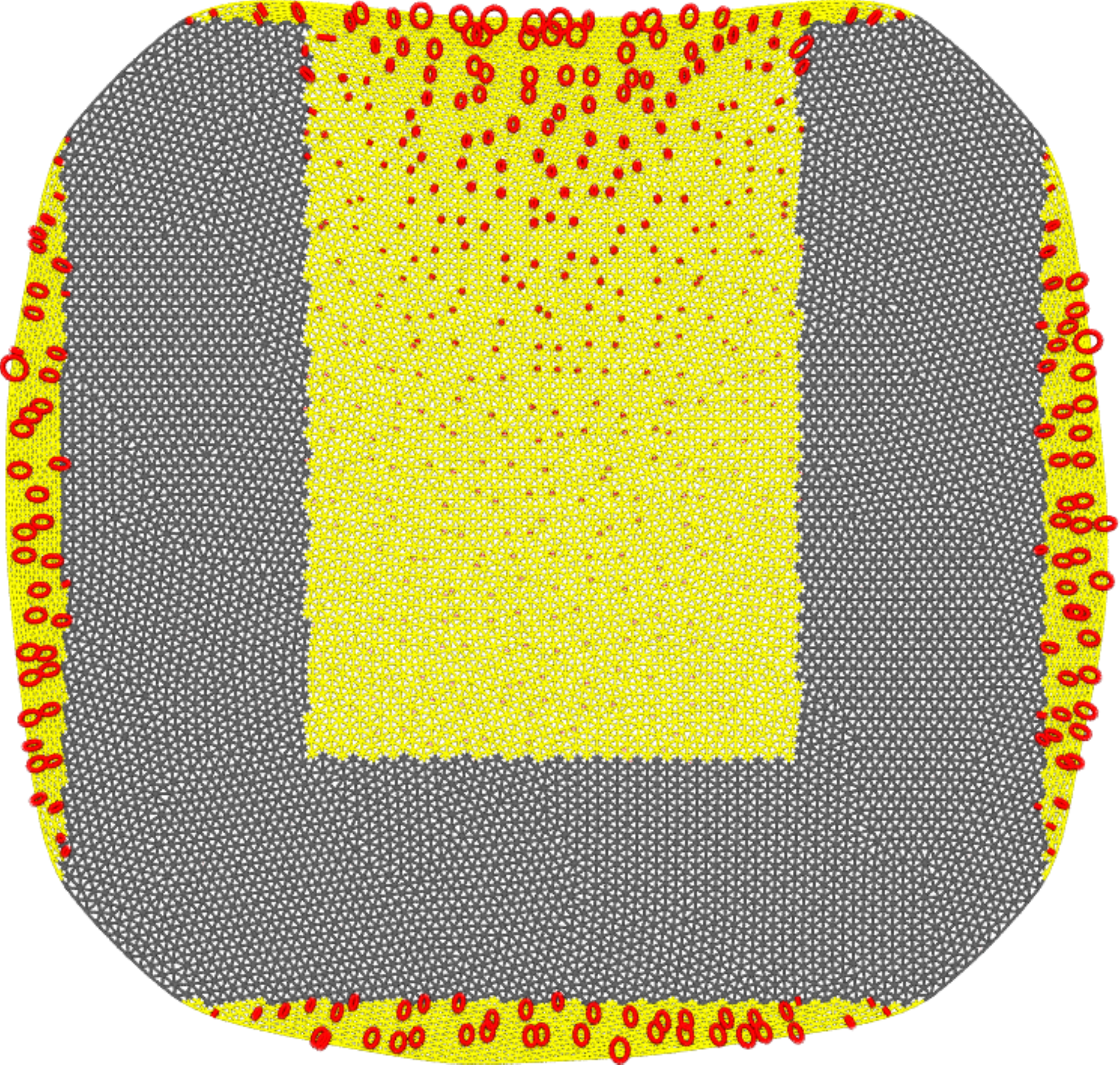}
\par
(b)
\end{center}%
\end{minipage}%
\begin{minipage}[t]{0.32\textwidth}%
\begin{center}
\includegraphics[scale=0.21]{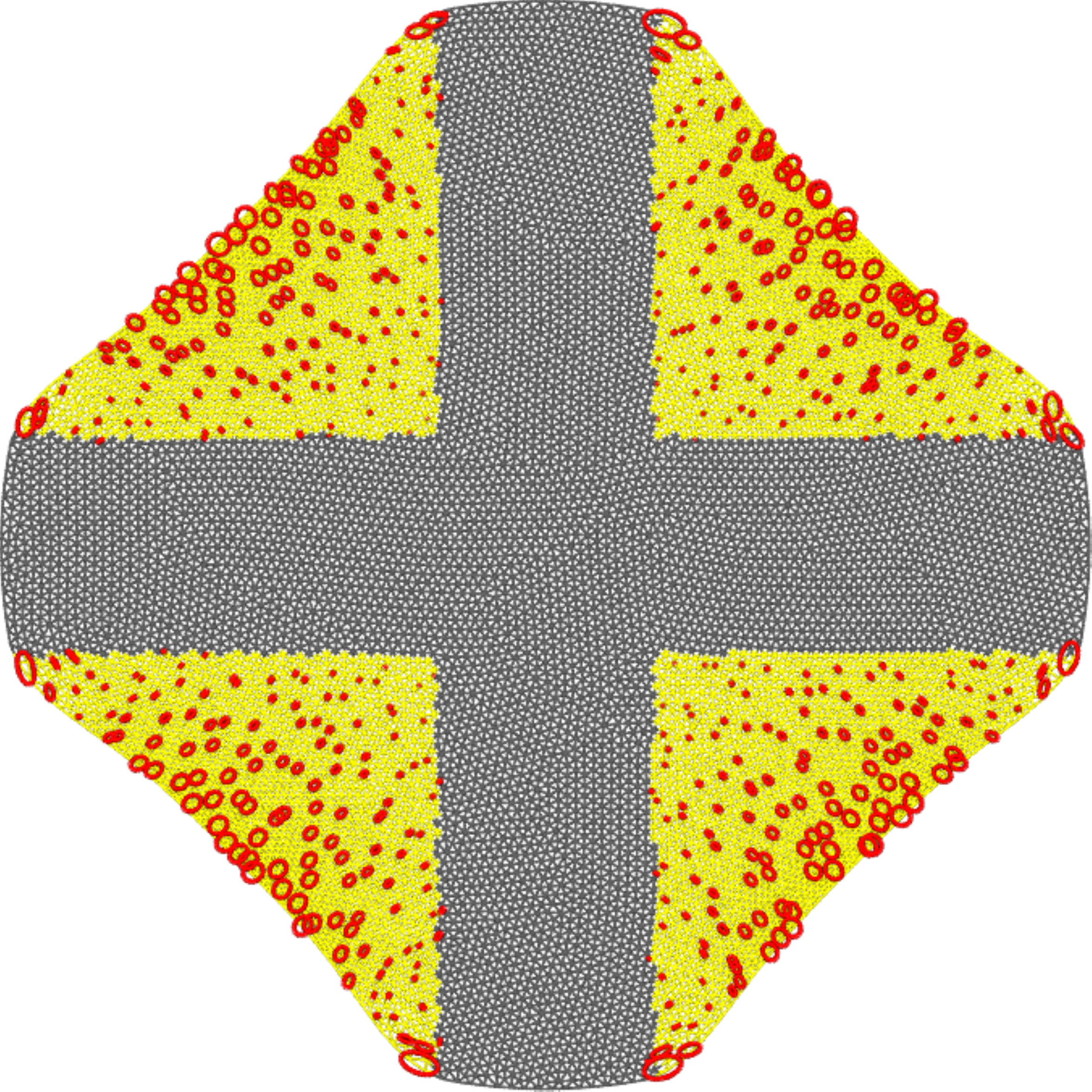}
\par
(c)
\end{center}%
\end{minipage}
\caption{Stress profile (Lam\'e's ellipses, shown in red) at representative sets of points of contracted cells for (a) $V$-pattern with $\kappa=0.16\ Y\ell^{3}$, (b) $U$-pattern with $\kappa=0.24\ Y\ell^{3}$, and (c) cross pattern with $\kappa=0.32\ Y\ell^{3}$; $\sigma_{a}=10Y$ for all patterns. Note that only the  elastic part of the stress tensor is shown. The active component, $\sigma_a$, has been removed, since it is isotropic and much larger in magnitude than the elastic component of the  stress tensor. Length of the ellipse axes is proportional to the two eigenvalues of the stress tensor, $\sigma_{min}$ and $\sigma_{max}$. The stress is not uniform but largest around cell's perimeter and gradually falls off toward its interior. For clarity, the Lam\'e's ellipses are computed only for a  subset of all triangles selected from the non-regular triangulation and their sizes are scaled by a factor of $0.5$.  }
\label{fig:stress}
\end{figure*}

Experiments probing the distribution of traction force density $\vec{T}(\vec{r})$  exerted by cells adhering to soft substrates  consistently show that such stresses are concentrated at the cell edges, and strongest in region of high cell curvature. Force balance requires $T_\alpha=\partial_\beta\sigma_{\alpha\beta}$, where $\sigma_{\alpha\beta}$ is the two-dimensional stress tensor of the bulk cellular material,  given by,
\begin{eqnarray}
\sigma_{\alpha\beta}&=&\frac{\partial\left(E_{el}+E_{active}\right)}{\partial u_{\alpha\beta}}=\sigma_{\alpha\beta}^{el}+\sigma_{a}\delta_{\alpha\beta}\nonumber\\
&=&\frac{Eh}{\left(1+\nu\right)}\left(\frac{\nu}{1-\nu}\delta_{\alpha\beta}u_{\gamma\gamma}+u_{\alpha\beta}\right)+\sigma_{a}\delta_{\alpha\beta}\;.\label{eq:stress}
\end{eqnarray}
The distribution of such internal stresses can therefore be inferred experimentally from traction force microscopy measurements~\cite{tambe2011}. Internal stresses of adhering cells are  found to be concentrated at the cell's interior, with a maximum value proportional to the active cell contractility, here $\sigma_a$. To highlight the role of patterned adhesion on the spatial distribution of cellular stress, we display in Fig.~\ref{fig:stress} the spatial distribution of the so-called
Lam\'e's stress ellipses~\cite{Love1927} for the elastic part $\sigma^{el}_{\alpha\beta}$ of the stress tensor. The constant active contribution $\sigma_a$ has been subtracted out to highlight spatial variations. As a result, the displayed stress is largest at the cell edges. The Lam\'e's stress ellipses are obtained by  computing the elastic part of the two-dimensional stress tensor at a representative subset of triangles in the Delaunay grid. This is achieved by directly evaluating the expression in eqn~\eqref{eq:stress}, excluding the active term. We then compute the low and the high eigenvalues, $\sigma_{min}$ and $\sigma_{max}$, respectively, of the elastic stress tensor of a given triangle. Note that since the stress tensor is symmetric, its eigenvalues are always real. The length of the major and minor semi-axes of each ellipse are then given by $\sigma_{max}$ and $\sigma_{min}$, respectively, whereas the orientation of the ellipse axes is given by the directions of the corresponding 
eigenvectors. As expected, and consistent with experiments~\cite{Thery2010}, elastic stresses are concentrated at the free boundaries of the adherent cell. Boundary stresses along free edges connecting two adhesion points are directed normal to the edge whereas they are oriented along the edge near the adhesion points. This is most evident in the cross-shaped pattern, Fig.~\ref{fig:stress}(c).  The large stresses in the convex regions of the cell spilling outside straight portions of the pinning regions (see, for instance, the V-shape pattern, Fig.~\ref{fig:stress}(a)) are largely an artifact of our model. They arise because we have introduced excluded volume interactions  to prevent self-intersections of the triangulation. In other words, we assign a hard core radius of $0.25\ell$ to each vertex of the triangulation, such that no two vertices can came closer than $0.5\ell$ from each other. Once this limit has been reached, the excluded volume prevents further collapse of the cell, thus accounting for the presence of  a sizeable portion of the cell that extends outside the pinning region. While steric effects are present \textit{in vivo} and may describe for instance the role of structural elements capable of carrying compressive loads, such as microtubules, cells on synthetic substrates generally almost completely conform to the micropattern by changing their thickness. This is not possible in our strictly two-dimensional model. As a result, the model captures well the behavior of ``free''  cell edges spanning two adhesion points, but has  limitations for describing the behavior of cell boundaries  along straight pinning regions.
 
\section{Boundary shapes}
\label{sec:Boundary_shapes}
\subsection{Strong pinning at adhesions}
\begin{figure*}
\centering
\includegraphics[width=0.85\textwidth]{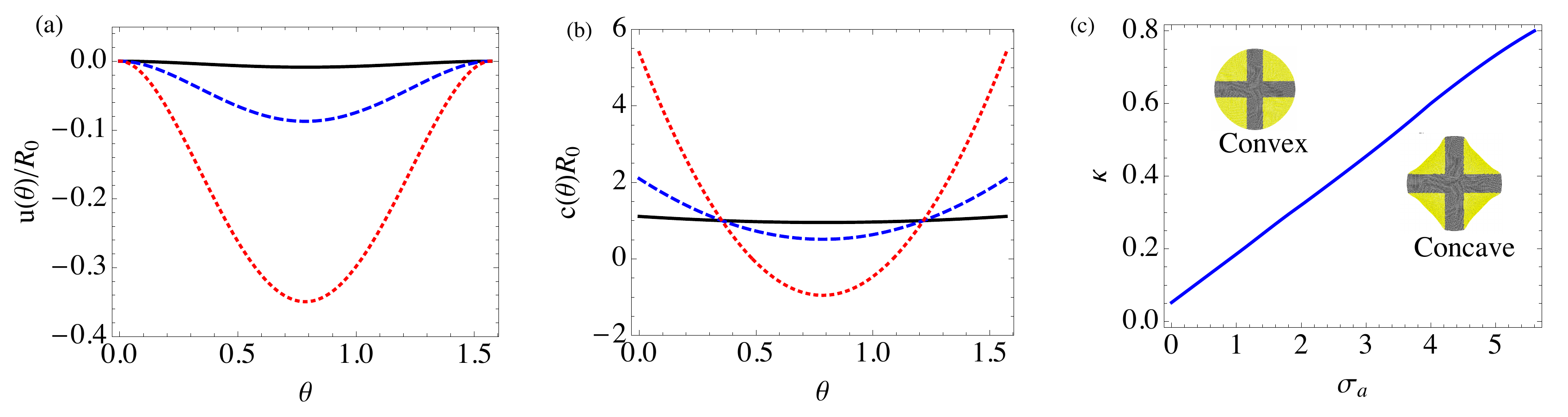}
\caption{\label{fig:theory}Curvature, deformation and phase boundary for a pinned contractile string. (a) Radial displacement $u_r$ in units of $R_0$ and (b) curvature profile $c(\theta)$ in units of $1/R_0$ for $\sigma_aR_0^3/\kappa=1$ (solid, black), $\sigma_aR_0^3/\kappa=10$ (dashed, blue) and $\sigma_aR_0^3/\kappa=40$ (dotted, red), where $0<\theta<\pi/2$, corresponding to a cross pattern of width zero. (c) Shape phase diagram for the contractile string pinned to a cross micropattern of width $w$, obtained from the solution to eqn~\eqref{eq:profile}. Bending rigidity $\kappa$ and contractile stress $\sigma_a$ are given in units of $Y\ell^3$ and $Y$ respectively, corresponding to the parameters in Fig.~\ref{fig:phase-diagram}, where $R_0/\ell=50$, $w/\ell=20$, and $\nu=1/3$.}
\end{figure*}
\label{sub:String_pinning}
For small deformations about an initially circular configuration of radius $R_0$, the cell boundary in the non-adhesive region can be parametrized using polar coordinates as, $r(\theta)=R_0+u_r(\theta)$, where $u_r$ is the radial component of the displacement field at the cell boundary, $\vec{u}=(u_r,u_\theta)$. Thus, $\vec{u}$ is solely a function of the angular coordinate $\theta$. In mechanical equilibrium, boundary force balance along the normal and tangential directions requires
\begin{subequations}
\begin{gather}
2\kappa\frac{d^2 c}{ds^2} - \sigma_{ij}n_in_j=0 \;,\label{eq-normal}\\
\sigma_{ij}t_in_j=0\;,
\end{gather}
\end{subequations}
 where $s$ is the arc-length parameter and $\vec{n}$ and $\vec{t}$ are  unit vectors normal and tangent, respectively, to the unperturbed cell boundary. Tangential force-balance in polar coordinates reduces to $\sigma_{r\theta}=0$, which leads to the relation $u_\theta=\partial_\theta u_r$.
Thus, the normal component of the elastic stress is given by $\sigma_{rr}^{el}\simeq \lambda \partial_\theta u_\theta/R_0=\lambda\partial_\theta^2 u_r/R_0$, where $\lambda=Y\nu/(1-\nu^2)$ is the Lam\'e elastic constant. Furthermore, for small deformations $u_r$, the boundary curvature can be expanded as,
\begin{equation}\label{eq-curvature}
c(\theta)\simeq\frac{1}{R_0}-\frac{1}{R_0^2}(u_r + \partial_\theta^2 u_r) + \mathcal{O}(u_r^2)\;.
\end{equation}
Using eqns~\eqref{eq-normal}-\eqref{eq-curvature}, and letting $ds=R_0d\theta$, we obtain an equation for the boundary profile,
\begin{equation}\label{eq:profile}
\frac{2\kappa}{R_0^3}\left(\partial_\theta^2 \tilde{u} + \partial_\theta^4 \tilde{u}\right) + \sigma_a +\lambda\partial_\theta^2\tilde{u}=0\;,
\end{equation}
where $\tilde{u}=u_r/R_0$. Without loss of generality, we can consider solution in the interval $\theta\in[0,\phi]$, where $\phi$ is the angular width of the nonadherent region and depends on the geometry of the adhesion pattern. The boundary conditions for the case of strong pinning at adhesions are given by: $\tilde{u}(0)=\tilde{u}(\phi)=\partial_\theta\tilde{u}(0)=\partial_\theta\tilde{u}(\phi)=0$. The full solution of eqn~\eqref{eq:profile} is analytically tractable but cumbersome. We instead discuss the solutions in two limiting cases in terms of the dimensionless parameter $K=2\kappa/\lambda R_0^3$, reflecting the relative contributions of bending and bulk elasticity. This parameter can also be written as $K=(\xi/R_0)^3$ in terms of the ratio of a length scale $\xi=[2\kappa/\lambda]^{1/3}$ to the undeformed cell radius $R_0$. The length scale $\xi$ described the interplay between bulk elasticity and boundary tension in controlling the response of the cell. When $\xi\gg R_0$ (corresponding to $K\gg1$) 
the cel deformation is controlled by the cortical tension at the boundary and the curvature is given by
\begin{equation}
c(\theta)\simeq\frac{1}{R_0}\left[1+\frac{\sigma_aR_0^3}{4\kappa}\left(2+\theta^2-\theta\phi-\phi\cot{\frac{\phi}{2}}\right)\right]\;.
\end{equation}
Bulk elasticity drops out and the behavior is controlled by the ratio $\sigma_aR_0^3/\kappa$ of contractility to bending rigidity.
The curvature has a minimum at the center of the nonadherent segment. Thus, as one increases contractility $\sigma_a$, a region of negative curvature develops near $\theta=\phi/2$, which grows upon increasing $\sigma_a$ until convexity is retained within a small neighborhood of the adhesion patch. The onset of concavity is thus given by the condition of reality to the solution of $c(\theta)=0$, which gives the condition,
\begin{equation}\label{eq:phase}
\sigma_a>\frac{2\kappa}{R_0^3}\left(\frac{1}{\phi^2/8 + (\phi/2)\cot{\frac{\phi}{2}}-1}\right)\;.
\end{equation}
Since concave shapes are commonly observed in experiments~\cite{Thery2006}, we now turn to estimate the critical value of $\sigma_a$ as predicted by our model in order to compare it with experimentally reported values for $\sigma_a$. The bending rigidity of cortical stress fibers can be estimated as $\kappa\sim \frac{\pi}{4} E_{act} r_s^{4}$, where $E_{act}$ is the Young's modulus of actin and $r_s$ is the typical radius of the stress fibers. Using $E_{act}\simeq 2.6$ GPa~\cite{Gittes1993} and $r_s\sim 0.1\ \mu$m, we get $\kappa\sim 2.0\times10^{-19}$ Nm$^2$. Using this value in eqn~\eqref{eq:phase} for $\phi\sim2\pi/3$, corresponding to a thin V-pattern, we get value for the critical $\sigma_a\sim$2.7 nN/$\mu$m. This is indeed the order of magnitude value for active stress or surface tension reported in experiments for adherent epithelial cells on continuous elastic substrates or endothelial cells on microposts~\cite{Bischofs2009,Mertz2012}.
In the opposite limit of $K\ll 1$ the deformation is controlled by bulk elasticity and the curvature is given by
\begin{equation}
c(\theta)\simeq\frac{1}{R_0}\left[1+\frac{\sigma_a}{2\lambda}\left(2+\theta^2-\theta\phi\right)\right]
\end{equation}
The condition of concavity is given by $\sigma_a>\lambda\left(\frac{1}{\phi^2/8-1}\right)$.
In the case when $\xi$ is comparable to $R_0$, a simpler solution of eqn~\eqref{eq:profile} can be obtained by neglecting the fourth order gradient term and also the derivative boundary conditions. The crossover to concave profiles can be approximated by the following interpolating form between the two limiting cases,
 \begin{equation}
  \sigma_a>\left(\lambda+\frac{2\kappa}{R_0^3}\right)\left[\frac{1}{\phi^2/8 + (\phi/2)\cot{\frac{\phi}{2}}-1}\right]\;.
 \end{equation}
 In the general case of eqn~\eqref{eq:profile}, the solution for curvature and the radial displacement is given in Fig.~\ref{fig:theory}a,b for three different values of $\sigma_aR_0^3/\kappa$ that compares the relative strengths of contractility to bending deformations. Furthermore, to compare numerically with the simulation results for the shape phase diagram of the adherent cell, we show the concave-convex phase boundary in $\sigma_a-\kappa$ plane in Fig.~\ref{fig:theory}c, for a cross-shaped micropattern using the same parameters as used in Fig.~\ref{fig:phase-diagram}. The resultant phase diagram is in good order-of-magnitude agreement with the simulation results, and the discrepancy in numerical values possibly arise from neglecting non-local bulk elasticity in the theoretical analysis.
 \vspace{0.4in}
\subsection{Soft pinning}
\label{sub:Soft_pinning}
We now  consider the case of soft pinning, where the free cell boundary is anchored to soft springs at the adhesion sites. Equation~\eqref{eq:profile} is now solved with the boundary conditions $\tilde{u}(0)=\tilde{u}(\phi)=\delta$ and $\partial_\theta\tilde{u}(0)=\partial_\theta\tilde{u}(\phi)=0$, where we have introduced an unknown displacement $\delta$ of the ends of the segment, which can be self consistently determined by minimizing the total energy of the deformed configuration with respect to $\delta$. For simplicity we ignore bulk elasticity and consider the limit $K\gg1$. The total energy of the deformed configuration is then given by $U=\kappa R_0\int_0^\phi d\theta c(\theta)^2 + k_s\delta^2$, where $k_s=\Gamma A_f$ and $A_f$ is the cross-sectional area of focal adhesions. Note that the contribution due to contractility vanishes in the final energy due to the derivative boundary conditions on $\tilde{u}$. The onset of concavity now depends on the substrate stiffness $k_s$ and the condition for 
convex-concave transition is given by
\begin{equation}
\sigma_a>\frac{4\kappa}{R_0^3}\left[\frac{k_s}{\kappa\phi^3/12R_0^3+ k_s\left(\phi^2/4 + \phi\cot{\frac{\phi}{2}}-2\right)}\right]\;.
\end{equation}
Thus, stiffer adhesions with $k_s R_0^3\gg\kappa$, promote concavity transition at a much higher value of contractility. It is favorable for a cell to invaginate at the free edges if the anchoring at adhesions is softer than the effective bending stiffness $\kappa/R_0^3$.

\section{Concluding Remarks}
\label{sec:Discusson}
Using a simple continuum model, coupling bulk and contour mechanics, we investigate the equilibrium shapes and stresses of adherent cells on substrates with various adhesion patterns. A continuum model without contour elasticity have been studied previously by two of us on convex patterns~\cite{Banerjee2011}, which was successful in capturing distribution of traction and cellular stresses and their dependence on substrate physical properties~\cite{Banerjee2012}. Here we focus on the shape and geometry induced stresses of non-adherent cell edges on concave micropatterns. We demonstrate numerically and analytically that the curvature of the non-adherent cell boundary can undergo a shape transition from convex to concave morphology, controlled by the interplay of contractility and bending rigidity. Stiff boundaries with low contractility relax to convex shapes, whereas at higher values of contractility, non-adherent cell edges attain a concave morphology. Previous work has shown that contractile cable network models are capable of reproducing the invaginated circular arc morphology of cell edges connecting strongly adhering sites~\cite{Torres2012}. Here we demonstrate that simple \emph{continuum} whole-cell models can also predict qualitatively cell shape and the transition between convex and concave cell edges, provided a bending rigidity describing cortical tension is included. For parameters realistic to experiments (see section 4.1) our model suggest that cells prefer to invaginate at their free edges, such that the effective boundary stiffness on non-adhesive zones are softer than myosin induced contractile stresses.  Images of actin from experiments on concave micropatterns do indeed show the formation of long and thin stress fibers that are invaginated on non-adherent edges~\cite{Thery2006}, indicating a softer cortical rigidity. In addition, elastic stresses are found to be higher along the free cell boundaries than in the neighborhood of adhesions, since in the absence of mechanotransduction cellular forces along free edges are not  shared by the substrate. Previous theoretical study with only contour elasticity indicated that substrate stiffness and contractility can cooperatively control cell morphology and induce hysteresis at the onset of convex-concave transition~\cite{Banerjee2013b}. Here we show that even in the presence of rigid adhesions, cell shape can be controlled by regulating the cortical bending rigidity and contractility.   Bending rigidity can be experimentally controlled by regulating the amount of actin cross-linking proteins that can impact stress fiber thickness and rigidity, whereas myosin based contractility can be perturbed using the conventional inhibitor Blebbistatin. 
One  limitation of our model is that it is strictly two-dimensional and does not allow for changes in the cell thickness.  Due to the presence of steric interactions in the finite element simulations, the cell edges on flat adhesive segments do not fully relax to the flat morphology, but maintain a convex shape. This is in contrast to real cells that contract to adjust to the shape of the micropattern. A fully three-dimensional model can overcome this difficulty, and is a natural extension of our present work.

\section{Appendix : Simulation Details}
\subsection{Discrete Model}
The discrete version of the elastic and
active contraction energies can be expressed as a sum over triangles of the triangulation,
\begin{equation}
E_{el}=\sum_{T}\left\{ \frac{Eh}{8\left(1+\nu\right)}\left(\frac{\nu}{1-\nu}\left(\mathrm{Tr}\hat{F}_{T}\right)^{2}+\mathrm{Tr}\left(\hat{F}_{T}\right)^{2}\right)\right\} A_{T}\;,\label{eq:el_eng_disc}
\end{equation}
\begin{equation}
E_{active}=\sigma_{a}\sum_{T}A_T\mathrm{Tr}\hat{F}_{T}\;,
\end{equation}
where matrix $\hat{F}=\hat{g}^{-1}\hat{G}-\hat{I}$, with $\hat{g}$ ($\hat{G}$)
being discrete metric tensor of the reference (deformed) configuration.
$A_{T}=\frac{1}{2}\left|\vec{a}\times\vec{b}\right|$ is the area
of an undeformed triangle spanned by two vectors $\vec{a}$ and $\vec{b}$
pointing along its sides. The sum is carried over all triangles. Adhesion
energy is discretized as 
\begin{equation}
E_{adh}=\frac{1}{2}\sum_{i}\Gamma_{i}\left|\vec{r}_{i}-\vec{r}_{i}^{\left(o\right)}\right|^{2}A_{i},\label{eq:adh_disc}
\end{equation}
where $\Gamma_{i}=10^{6}$($0$) for grey (yellow) vertices in Fig.~\ref{fig:intitial},
$\vec{r}_{i}$($\vec{r}_{i}^{\left(0\right)})$ is the current (reference)
position of the vertex \emph{i}, and $A_{i}=\frac{1}{3}\sum_{T\in\Omega_{i}}A_{T}$
is the area associated to the vertex (i.e., a third of the sum of
areas of all triangles that share the vertex, so-called ``vertex
star''). Finally, following ref.~\citenum{Brakke1992}, the boundary
bending energy is discretized as
\begin{equation}
E_{bend}=4\kappa\sum_{i}\frac{1-\cos\left(\vartheta_{i}\right)}{s_{i}+s_{i+1}},\label{eq:bend_disc}
\end{equation}
where $\vartheta_{i}$ is the exterior angle at the boundary vertex
\emph{i}, $s_{i}$ and $s_{i+1}$ are lengths of two boundary edges
meeting at \emph{i}, and the sum is carried over all boundary vertices. 

\subsection{Monte Carlo Sweeps}
\label{sub:app_MC}
A Monte Carlo sweep consist of an attempted move for each vertex.
A randomly selected vertex is displaced by $\Delta\vec{r}$
where components of $\Delta\vec{r}$ are chosen at random with an
equal probability from an interval $\left[-0.01\ell,0.01\ell\right]$. Moves were accepted according to the Metropolis rules. Minimum energy configuration is obtained using simulated annealing with linear cooling
protocol. Typically minimum energy configurations were reached to a satisfactory precision within $10^{6}$ Monte Carlo sweeps.

\bibliography{cell-micropattern}

\providecommand*{\mcitethebibliography}{\thebibliography}
\csname @ifundefined\endcsname{endmcitethebibliography}
{\let\endmcitethebibliography\endthebibliography}{}
\begin{mcitethebibliography}{31}
\providecommand*{\natexlab}[1]{#1}
\providecommand*{\mciteSetBstSublistMode}[1]{}
\providecommand*{\mciteSetBstMaxWidthForm}[2]{}
\providecommand*{\mciteBstWouldAddEndPuncttrue}
  {\def\EndOfBibitem{\unskip.}}
\providecommand*{\mciteBstWouldAddEndPunctfalse}
  {\let\EndOfBibitem\relax}
\providecommand*{\mciteSetBstMidEndSepPunct}[3]{}
\providecommand*{\mciteSetBstSublistLabelBeginEnd}[3]{}
\providecommand*{\EndOfBibitem}{}
\mciteSetBstSublistMode{f}
\mciteSetBstMaxWidthForm{subitem}
{(\emph{\alph{mcitesubitemcount}})}
\mciteSetBstSublistLabelBeginEnd{\mcitemaxwidthsubitemform\space}
{\relax}{\relax}

\bibitem[Schwarz and Gardel(2012)]{Schwarz2012}
U.~S. Schwarz and M.~L. Gardel, \emph{Journal of cell science}, 2012,
  \textbf{125}, 3051--3060\relax
\mciteBstWouldAddEndPuncttrue
\mciteSetBstMidEndSepPunct{\mcitedefaultmidpunct}
{\mcitedefaultendpunct}{\mcitedefaultseppunct}\relax
\EndOfBibitem
\bibitem[Discher \emph{et~al.}(2005)Discher, Janmey, and Wang]{Discher2005}
D.~Discher, P.~Janmey and Y.~Wang, \emph{Science}, 2005, \textbf{310},
  1139--1143\relax
\mciteBstWouldAddEndPuncttrue
\mciteSetBstMidEndSepPunct{\mcitedefaultmidpunct}
{\mcitedefaultendpunct}{\mcitedefaultseppunct}\relax
\EndOfBibitem
\bibitem[Yeung \emph{et~al.}(2005)Yeung, Georges, Flanagan, Marg, Ortiz,
  Funaki, Zahir, Ming, Weaver, and Janmey]{Yeung2005}
T.~Yeung, P.~Georges, L.~Flanagan, B.~Marg, M.~Ortiz, M.~Funaki, N.~Zahir,
  W.~Ming, V.~Weaver and P.~Janmey, \emph{Cell Motil Cytoskel}, 2005,
  \textbf{60}, 24--34\relax
\mciteBstWouldAddEndPuncttrue
\mciteSetBstMidEndSepPunct{\mcitedefaultmidpunct}
{\mcitedefaultendpunct}{\mcitedefaultseppunct}\relax
\EndOfBibitem
\bibitem[Lo \emph{et~al.}(2000)Lo, Wang, Dembo, and Wang]{Lo2000}
C.-M. Lo, H.-B. Wang, M.~Dembo and Y.-l. Wang, \emph{Biophysical journal},
  2000, \textbf{79}, 144--152\relax
\mciteBstWouldAddEndPuncttrue
\mciteSetBstMidEndSepPunct{\mcitedefaultmidpunct}
{\mcitedefaultendpunct}{\mcitedefaultseppunct}\relax
\EndOfBibitem
\bibitem[Ghibaudo \emph{et~al.}(2008)Ghibaudo, Saez, Trichet, Xayaphoummine,
  Browaeys, Silberzan, Buguin, and Ladoux]{Ghibaudo2008}
M.~Ghibaudo, A.~Saez, L.~Trichet, A.~Xayaphoummine, J.~Browaeys, P.~Silberzan,
  A.~Buguin and B.~Ladoux, \emph{Soft Matter}, 2008, \textbf{4},
  1836--1843\relax
\mciteBstWouldAddEndPuncttrue
\mciteSetBstMidEndSepPunct{\mcitedefaultmidpunct}
{\mcitedefaultendpunct}{\mcitedefaultseppunct}\relax
\EndOfBibitem
\bibitem[Chopra \emph{et~al.}(2011)Chopra, Tabdanov, Patel, Janmey, and
  Kresh]{Chopra2011}
A.~Chopra, E.~Tabdanov, H.~Patel, P.~Janmey and J.~Kresh, \emph{Am J
  Physiol-Heart C}, 2011, \textbf{300}, H1252--H1266\relax
\mciteBstWouldAddEndPuncttrue
\mciteSetBstMidEndSepPunct{\mcitedefaultmidpunct}
{\mcitedefaultendpunct}{\mcitedefaultseppunct}\relax
\EndOfBibitem
\bibitem[Th{\'e}ry(2010)]{Thery2010}
M.~Th{\'e}ry, \emph{Journal of Cell Science}, 2010, \textbf{123},
  4201--4213\relax
\mciteBstWouldAddEndPuncttrue
\mciteSetBstMidEndSepPunct{\mcitedefaultmidpunct}
{\mcitedefaultendpunct}{\mcitedefaultseppunct}\relax
\EndOfBibitem
\bibitem[Chen \emph{et~al.}(1997)Chen, Mrksich, Huang, Whitesides, and
  Ingber]{Chen1997}
C.~Chen, M.~Mrksich, S.~Huang, G.~Whitesides and D.~Ingber, \emph{Science},
  1997, \textbf{276}, 1425--1428\relax
\mciteBstWouldAddEndPuncttrue
\mciteSetBstMidEndSepPunct{\mcitedefaultmidpunct}
{\mcitedefaultendpunct}{\mcitedefaultseppunct}\relax
\EndOfBibitem
\bibitem[Th{\'e}ry \emph{et~al.}(2006)Th{\'e}ry, P{\'e}pin, Dressaire, Chen,
  and Bornens]{Thery2006}
M.~Th{\'e}ry, A.~P{\'e}pin, E.~Dressaire, Y.~Chen and M.~Bornens, \emph{Cell
  Motil Cytoskel}, 2006, \textbf{63}, 341--355\relax
\mciteBstWouldAddEndPuncttrue
\mciteSetBstMidEndSepPunct{\mcitedefaultmidpunct}
{\mcitedefaultendpunct}{\mcitedefaultseppunct}\relax
\EndOfBibitem
\bibitem[Roca-Cusachs \emph{et~al.}(2008)Roca-Cusachs, Alcaraz, Sunyer,
  Samitier, Farr{\'e}, and Navajas]{Roca2008}
P.~Roca-Cusachs, J.~Alcaraz, R.~Sunyer, J.~Samitier, R.~Farr{\'e} and
  D.~Navajas, \emph{Biophysical journal}, 2008, \textbf{94}, 4984--4995\relax
\mciteBstWouldAddEndPuncttrue
\mciteSetBstMidEndSepPunct{\mcitedefaultmidpunct}
{\mcitedefaultendpunct}{\mcitedefaultseppunct}\relax
\EndOfBibitem
\bibitem[Rape \emph{et~al.}(2011)Rape, Guo, and Wang]{Rape2011}
A.~D. Rape, W.-h. Guo and Y.-l. Wang, \emph{Biomaterials}, 2011, \textbf{32},
  2043--2051\relax
\mciteBstWouldAddEndPuncttrue
\mciteSetBstMidEndSepPunct{\mcitedefaultmidpunct}
{\mcitedefaultendpunct}{\mcitedefaultseppunct}\relax
\EndOfBibitem
\bibitem[Bar-Ziv \emph{et~al.}(1999)Bar-Ziv, Tlusty, Moses, Safran, and
  Bershadsky]{Barziv1999}
R.~Bar-Ziv, T.~Tlusty, E.~Moses, S.~Safran and A.~Bershadsky, \emph{Proc Natl
  Acad Sci USA}, 1999, \textbf{96}, 10140--10145\relax
\mciteBstWouldAddEndPuncttrue
\mciteSetBstMidEndSepPunct{\mcitedefaultmidpunct}
{\mcitedefaultendpunct}{\mcitedefaultseppunct}\relax
\EndOfBibitem
\bibitem[Schwarz and Safran(2013)]{Schwarz2013}
U.~S. Schwarz and S.~A. Safran, \emph{Reviews of Modern Physics}, 2013,
  \textbf{85}, 1327\relax
\mciteBstWouldAddEndPuncttrue
\mciteSetBstMidEndSepPunct{\mcitedefaultmidpunct}
{\mcitedefaultendpunct}{\mcitedefaultseppunct}\relax
\EndOfBibitem
\bibitem[Bischofs \emph{et~al.}(2009)Bischofs, Schmidt, and
  Schwarz]{Bischofs2009}
I.~Bischofs, S.~Schmidt and U.~Schwarz, \emph{Phys Rev Lett}, 2009,
  \textbf{103}, 48101\relax
\mciteBstWouldAddEndPuncttrue
\mciteSetBstMidEndSepPunct{\mcitedefaultmidpunct}
{\mcitedefaultendpunct}{\mcitedefaultseppunct}\relax
\EndOfBibitem
\bibitem[Banerjee and Giomi(2013)]{Banerjee2013b}
S.~Banerjee and L.~Giomi, \emph{Soft Matter}, 2013, \textbf{9},
  5251--5260\relax
\mciteBstWouldAddEndPuncttrue
\mciteSetBstMidEndSepPunct{\mcitedefaultmidpunct}
{\mcitedefaultendpunct}{\mcitedefaultseppunct}\relax
\EndOfBibitem
\bibitem[Banerjee and Marchetti(2011)]{Banerjee2011}
S.~Banerjee and M.~C. Marchetti, \emph{EPL (Europhysics Letters)}, 2011,
  \textbf{96}, 28003\relax
\mciteBstWouldAddEndPuncttrue
\mciteSetBstMidEndSepPunct{\mcitedefaultmidpunct}
{\mcitedefaultendpunct}{\mcitedefaultseppunct}\relax
\EndOfBibitem
\bibitem[Edwards and Schwarz(2011)]{Edwards2011}
C.~M. Edwards and U.~S. Schwarz, \emph{Physical Review Letters}, 2011,
  \textbf{107}, 128101\relax
\mciteBstWouldAddEndPuncttrue
\mciteSetBstMidEndSepPunct{\mcitedefaultmidpunct}
{\mcitedefaultendpunct}{\mcitedefaultseppunct}\relax
\EndOfBibitem
\bibitem[Pathak \emph{et~al.}(2008)Pathak, Deshpande, McMeeking, and
  Evans]{Pathak2008}
A.~Pathak, V.~S. Deshpande, R.~M. McMeeking and A.~G. Evans, \emph{Journal of
  The Royal Society Interface}, 2008, \textbf{5}, 507--524\relax
\mciteBstWouldAddEndPuncttrue
\mciteSetBstMidEndSepPunct{\mcitedefaultmidpunct}
{\mcitedefaultendpunct}{\mcitedefaultseppunct}\relax
\EndOfBibitem
\bibitem[Banerjee and Marchetti(2013)]{Banerjee2013a}
S.~Banerjee and M.~C. Marchetti, \emph{New Journal of Physics}, 2013,
  \textbf{15}, 035015\relax
\mciteBstWouldAddEndPuncttrue
\mciteSetBstMidEndSepPunct{\mcitedefaultmidpunct}
{\mcitedefaultendpunct}{\mcitedefaultseppunct}\relax
\EndOfBibitem
\bibitem[Vianay \emph{et~al.}(2010)Vianay, K{\"a}fer, Planus, Block, Graner,
  and Guillou]{Vianay2010}
B.~Vianay, J.~K{\"a}fer, E.~Planus, M.~Block, F.~Graner and H.~Guillou,
  \emph{Physical review letters}, 2010, \textbf{105}, 128101\relax
\mciteBstWouldAddEndPuncttrue
\mciteSetBstMidEndSepPunct{\mcitedefaultmidpunct}
{\mcitedefaultendpunct}{\mcitedefaultseppunct}\relax
\EndOfBibitem
\bibitem[Torres \emph{et~al.}(2012)Torres, Bischofs, and Schwarz]{Torres2012}
P.~G. Torres, I.~Bischofs and U.~Schwarz, \emph{Physical Review E}, 2012,
  \textbf{85}, 011913\relax
\mciteBstWouldAddEndPuncttrue
\mciteSetBstMidEndSepPunct{\mcitedefaultmidpunct}
{\mcitedefaultendpunct}{\mcitedefaultseppunct}\relax
\EndOfBibitem
\bibitem[Heidemann and Wirtz(2004)]{Heidemann2004}
S.~R. Heidemann and D.~Wirtz, \emph{Trends in cell biology}, 2004, \textbf{14},
  160--166\relax
\mciteBstWouldAddEndPuncttrue
\mciteSetBstMidEndSepPunct{\mcitedefaultmidpunct}
{\mcitedefaultendpunct}{\mcitedefaultseppunct}\relax
\EndOfBibitem
\bibitem[Audoly and Pomeau(2010)]{audoly2010}
B.~Audoly and Y.~Pomeau, \emph{Elasticity and geometry: from hair curls to the
  non-linear response of shells}, Oxford University Press Oxford, 2010\relax
\mciteBstWouldAddEndPuncttrue
\mciteSetBstMidEndSepPunct{\mcitedefaultmidpunct}
{\mcitedefaultendpunct}{\mcitedefaultseppunct}\relax
\EndOfBibitem
\bibitem[Gardel \emph{et~al.}(2004)Gardel, Shin, MacKintosh, Mahadevan,
  Matsudaira, and Weitz]{Gardel2004}
M.~Gardel, J.~Shin, F.~MacKintosh, L.~Mahadevan, P.~Matsudaira and D.~Weitz,
  \emph{Science}, 2004, \textbf{304}, 1301--1305\relax
\mciteBstWouldAddEndPuncttrue
\mciteSetBstMidEndSepPunct{\mcitedefaultmidpunct}
{\mcitedefaultendpunct}{\mcitedefaultseppunct}\relax
\EndOfBibitem
\bibitem[Mertz \emph{et~al.}(2012)Mertz, Banerjee, Che, German, Xu, Hyland,
  Marchetti, Horsley, and Dufresne]{Mertz2012}
A.~Mertz, S.~Banerjee, Y.~Che, G.~German, Y.~Xu, C.~Hyland, M.~Marchetti,
  V.~Horsley and E.~Dufresne, \emph{Phys Rev Lett}, 2012, \textbf{108},
  198101\relax
\mciteBstWouldAddEndPuncttrue
\mciteSetBstMidEndSepPunct{\mcitedefaultmidpunct}
{\mcitedefaultendpunct}{\mcitedefaultseppunct}\relax
\EndOfBibitem
\bibitem[Banerjee and Marchetti(2012)]{Banerjee2012}
S.~Banerjee and M.~C. Marchetti, \emph{Phys Rev Lett}, 2012, \textbf{109},
  108101\relax
\mciteBstWouldAddEndPuncttrue
\mciteSetBstMidEndSepPunct{\mcitedefaultmidpunct}
{\mcitedefaultendpunct}{\mcitedefaultseppunct}\relax
\EndOfBibitem
\bibitem[Weeks \emph{et~al.}(1971)Weeks, Chandler, and Andersen]{Weeks1971}
J.~D. Weeks, D.~Chandler and H.~C. Andersen, \emph{The Journal of chemical
  physics}, 1971, \textbf{54}, 5237\relax
\mciteBstWouldAddEndPuncttrue
\mciteSetBstMidEndSepPunct{\mcitedefaultmidpunct}
{\mcitedefaultendpunct}{\mcitedefaultseppunct}\relax
\EndOfBibitem
\bibitem[Tambe \emph{et~al.}(2011)Tambe, Hardin, Angelini, Rajendran, Park,
  Serra-Picamal, Zhou, Zaman, Butler, Weitz,\emph{et~al.}]{tambe2011}
D.~T. Tambe, C.~C. Hardin, T.~E. Angelini, K.~Rajendran, C.~Y. Park,
  X.~Serra-Picamal, E.~H. Zhou, M.~H. Zaman, J.~P. Butler, D.~A. Weitz
  \emph{et~al.}, \emph{Nature Materials}, 2011, \textbf{10}, 469--475\relax
\mciteBstWouldAddEndPuncttrue
\mciteSetBstMidEndSepPunct{\mcitedefaultmidpunct}
{\mcitedefaultendpunct}{\mcitedefaultseppunct}\relax
\EndOfBibitem
\bibitem[Love(1927)]{Love1927}
A.~E.~H. Love, \emph{A {T}reatise on the {M}athematical {T}heory of
  {E}lasticity}, Cambridge University Press, Cambridge, 4th edn, 1927\relax
\mciteBstWouldAddEndPuncttrue
\mciteSetBstMidEndSepPunct{\mcitedefaultmidpunct}
{\mcitedefaultendpunct}{\mcitedefaultseppunct}\relax
\EndOfBibitem
\bibitem[Gittes \emph{et~al.}(1993)Gittes, Mickey, Nettleton, and
  Howard]{Gittes1993}
F.~Gittes, B.~Mickey, J.~Nettleton and J.~Howard, \emph{The Journal of cell
  biology}, 1993, \textbf{120}, 923--934\relax
\mciteBstWouldAddEndPuncttrue
\mciteSetBstMidEndSepPunct{\mcitedefaultmidpunct}
{\mcitedefaultendpunct}{\mcitedefaultseppunct}\relax
\EndOfBibitem
\bibitem[Brakke(1992)]{Brakke1992}
K.~A. Brakke, \emph{Experimental mathematics}, 1992, \textbf{1}, 141--165\relax
\mciteBstWouldAddEndPuncttrue
\mciteSetBstMidEndSepPunct{\mcitedefaultmidpunct}
{\mcitedefaultendpunct}{\mcitedefaultseppunct}\relax
\EndOfBibitem
\end{mcitethebibliography}

\end{document}